\documentclass[twocolumn,superscriptaddress,showpacs,preprintnumbers,amsmath,amssymb,floatfix]{revtex4}

\usepackage{graphicx,amssymb}

\begin{document}

\title{Clusters, columns, and lamellae -- minimum energy configurations in 
core softened potentials}

\author{Gernot J. Pauschenwein} 
\affiliation{Center for Computational Materials Science and Institut
f\"ur Theoretische Physik, Technische Universit\"at Wien, Wiedner
Hauptstra{\ss}e 8-10, A-1040 Wien, Austria} 

\author{Gerhard Kahl}
\affiliation{Center for Computational Materials Science and Institut
f\"ur Theoretische Physik, Technische Universit\"at Wien, Wiedner
Hauptstra{\ss}e 8-10, A-1040 Wien, Austria}

\date{\today}

\begin{abstract}
We give evidence that particles interacting via the simple, radially
symmetric square-shoulder potential can self-organize in highly
complex, low-symmetry lattices, forming thereby clusters, columns, or
lamellae; only at high pressure compact, high-symmetry structures are
observed. Our search for these ordered equilibrium structures is based
on ideas of genetic algorithms, a strategy that is characterized by a
high success rate. A simple mean-field type consideration complements
these findings and locates in a semi-quantitative way the cross-over
between the competing structures.
\end{abstract}


\maketitle

''One of the continuing scandals in the physical sciences is that it
remains in general impossible to predict the structure of even the
simplest crystalline solids from a knowledge of their chemical
composition'' \cite{Maddox88}. Even nowadays, twenty years later, this
statement is still valid and it applies equally well to the case that
the {\it physical} properties of the system, e.g., in terms of its
inter-particle potential, are known. Indeed, in many problems of hard
and soft condensed matter theory a powerful tool that is able to {\it
predict} the ordered equilibrium structures of a system in a reliable
way is badly missing. This applies in particular to soft matter
systems where particles are able to self-organize in a broad variety
of unexpected and often very exotic ordered structures: micellar and
inverse micellar structures \cite{Pie06,Gla07}, spirals \cite{Cam05},
chains and layers \cite{Gla07,Can06}, and cluster phases
\cite{Str04,Mla06} are a few examples.

During the past decades several strategies have been proposed to find
the energetically most favorable particle arrangements of a
system. Apart from conventional approaches that rely on intuition,
experience, or plausible arguments when selecting candidates for
ordered equilibrium structures, there are more sophisticated
approaches such as simulated annealing \cite{Dee89Pan90Boi95Schoen96},
basin hopping \cite{Goe04}, or meta-dynamics
\cite{Mar03Mar05Mar06}. However, all these strategies are affected by
different sorts of deficiencies which can significantly reduce their
success rates.

In recent years convincing evidence has been given that search
strategies based on ideas of genetic algorithms (GAs) are able to
provide a significant break-through to solve this problem. Generally
speaking, GAs are strategies that use key ideas of evolutionary
processes, such as survival of the fittest, recombination, or
mutation, to find optimal solutions for a problem \cite{Hol75}. The
wide spectrum of obviously successful applications in different fields
of condensed matter physics unambiguously demonstrates their
flexibility, reliability, and efficiency: among these are laser pulse
control \cite{Ass98}, protein folding \cite{Ste94}, or cluster
formation \cite{Dae95}. In contrast, attempts to apply GAs in the
search for ordered equilibrium structures in condensed matter theory
were realized considerably later. While the first application probably
dates back to 1999 \cite{Woo99} their widespread use in hard matter
theory was pioneered by Oganov and co-workers \cite{Oga06Oga08} only
recently, where they have become meanwhile a standard tool: a wide
spectrum of successful applications ranging from geophysical to
technologically relevant problems give evidence of the power and the
flexibility of this approach (for an overview see \cite{Oga06Oga08}).
In {\it soft} condensed matter theory the usage of these
search strategies is still in its infancy. First applications to find
minimum energy configurations (MECs) of soft systems have,
nevertheless, unambiguously documented the power of the algorithm:
successful examples are the identification of exotic lattice
structures and cluster phases for particular soft systems
\cite{Got04,Mla06}, or of complex, ordered arrangements of monolayers
of binary dipolar mixtures \cite{For08LoV}. All these investigations
mentioned above give evidence that GA-based search strategies have an
extremely high success rate.

In this contribution we consider a simple soft model system, i.e., a
square-shoulder system, and show that the GA is an efficient and
reliable tool to identify even highly complex MECs in soft matter
systems. We discover an overwhelming and undoubtedly unexpected wealth
of ordered MECs, which comprise clusters, columns, lamellae, as well
as compact structures. The simple shape of the potential allows an
easy geometric interpretation of the system's strategies to form MECs
in terms of overlapping shoulders, a nice feature that is not
available in systems with continuous potentials. We find evidence that
the success rate of our algorithm must be close to 100\%.

The square-shoulder system is the simplest representative in the class
of the so-called core-softened potentials (for references see
\cite{Yan06}).  Despite their innocently looking potentials these
systems are characterized by a host of surprising properties, such as
water-like anomalies \cite{Jag99,Yan06} or a rich wealth in the
occurring structures, investigated in detail in two-dimensional
systems \cite{For08LoV,Gla07}. The potential $\Phi(r)$ of the
square-shoulder system,
\begin{equation}
\Phi(r) = \left\{ \begin{array} {l@{~~~~~~~~}l} 
                  \infty & r \le \sigma \\ 
                  \epsilon & \sigma < r \le \lambda \sigma \\
                  0 & \lambda \sigma < r  \\
                             \end{array}
                             \right. ,
\end{equation}
consists of an impenetrable core of diameter $\sigma$ with
an adjacent, repulsive, step-shaped shoulder of height $\epsilon$ and
width $\lambda \sigma$. For the aims of the present contribution it represents the
'quintessential' test system \cite{Zih01}. But we also point out that
the square-shoulder system is not only of purely academic interest: it
represents a reasonable model system for colloidal particles with a
core-corona architecture, as has been given evidence for in
\cite{Nor05}. Further, we introduce the number-density $\varrho = N/V$,
$N$ and $V$ being the number of particles and the volume,
respectively.

The MECs for this system have been identified via a GA-based search
strategy in the NPT ensemble. For details of the encoding of the
individuals (= lattices) and of the reproduction mechanism we refer to
\cite{Got05Lik}. The quality of an individual ${\cal I}$ is measured
via the fitness function $f({\cal I})$, for which we haven chosen
$f({\cal I}) = \exp\{-[G({\cal I}) - G_0]/G_0\}$. Since our
investigations are restricted to the case $T=0$, the Gibbs free
energy, $G({\cal I})$, reduces at a given pressure $P$ to $G=U+PV$,
$U$ being the lattice sum; $G_0$ is the Gibbs free energy of a
reference structure. Significant extensions of the search strategy
were required due to the impenetrable core: as a consequence of the
highly stochastic character of GAs, the algorithm will propose with a
high probability lattices where the cores of the particles overlap:
such configurations are unphysical and thus useless. A more {\it
quantitative} investigation reveals that the physically relevant
lattices (characterized by non-overlapping cores of the particles)
populate only a highly porous subset of the search space
\cite{comment}.
 
To overcome this problem a suitable modification is urgently required,
which guarantees that unphysical configurations are excluded {\it a
priori} so that only lattices with non-overlapping cores are
created. Such a strategy has been developed and will be outlined
briefly. We start for simplicity with a simple lattice, described via
a set of linearly independent vectors ${\bf a}_1$, ${\bf a}_2$, and
${\bf a}_3$. In order to satisfy our expectations, they have to meet
several requirements. First, the vectors are chosen such that $|{\bf
a}_1| \le |{\bf a}_2| \le |{\bf a}_3|$, where $|{\bf a}_1|$ represents
the shortest possible distance between lattice sites. Then, ${\bf
a}_2$ is selected so that $|{\bf a}_2|$ is either equal to $|{\bf
a}_1|$ or represents the second smallest distance encountered between
two lattice points. Finally, the third vector, ${\bf a}_3$, is chosen
so that $|{\bf a}_3|$ satisfies a similar relation with respect to
$|{\bf a}_2|$. Of course, the choice of the $\{ {\bf a}_i \}$ is not
unique. If we are able to construct a lattice so that {\it a priori}
$|{\bf a}_1| \ge \sigma$, then it is guaranteed by construction that
the particles will not overlap. We have succeeded in developing a
formalism that is able to create vectors $\{ {\bf a}_i \}$, that
satisfy the above requirements. These constraints lead to inequalities
between the Cartesian components of the vectors; these lengthy
relations along with a detailed explanation of the algorithm will be
described elsewhere \cite{Pau08}. In non-simple lattices overlap can
also be caused by the basis particles. Let us assume that the
underlying {\it simple} lattice satisfies above requirements. Then we
calculate all the distances between the particles of this cell and all
particles located in the 26 neighboring cells. Let $l_0$ be the
smallest among these distances; if $l_0<|{\bf a}_1|$, then the vectors $\{
{\bf a}_i \}$ are scaled by a factor $|{\bf a}_1|/l_0$, which guarantees
that in the entire lattice no particle overlap will occur.

\begin{figure}
      \begin{center}
      \begin{minipage}[t]{8.5cm}
      \includegraphics[width=8.4cm, clip] {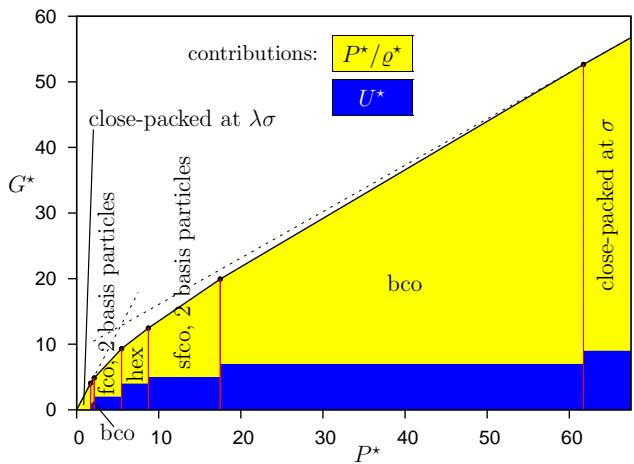}
      \end{minipage}
      \end{center}
\caption{(color online). $G^\star$ and $U^\star$ as functions of
  $P^\star$ for the square-shoulder system, $\lambda = 1.5$. Shaded
  areas represent contributions to $G^\star$ as labeled. Broken lines
  indicate low- and high-pressure limiting cases of MECs (see text).}
\label{fig:thd_1.5}
\end{figure}

Thermodynamic properties will be presented in standard reduced units:
$\varrho^\star=\varrho\sigma^3$, $P^\star = P \sigma^3/\epsilon$,
$U^\star = U/(N\epsilon)$, and $G^\star = G/(N\epsilon) = U^\star +
P^\star/\varrho^\star$. The simple shape of the potential allows us to
simplify the search considerably: since for a fixed particle
configuration $U^\star$ (counting the number of overlapping coronas)
is constant, $G^\star = U^\star + P^\star/\varrho^\star$ is a straight
line of slope $1/\varrho^\star$ in the $(G^\star, P^\star)$-diagram.
The limiting MECs at low and high pressure are easily identified as
close-packed spheres, either with diameter $\lambda \sigma$
(corresponding to a slope $1/\varrho^\star_{\rm min} = \lambda^3/\sqrt
2$) or with diameter $\sigma$ (corresponding to a slope
$1/\varrho^\star_{\rm max} = 1/\sqrt 2$) -- cf.\
Fig. \ref{fig:thd_1.5}. Any other MEC occurring in this system is
characterized by a line of slope $1/\varrho^\star$, satisfying
$1/\varrho^\star_{\rm min} > 1/\varrho^\star > 1/\varrho^\star_{\rm
max}$. Thus $G^\star = G^\star (P^\star)$ is a sequence of straight
lines. In a first step we determine in the $(G^\star,
P^\star)$-diagram the intersection point of the two limiting straight
lines mentioned above and launch for this $P^\star$-value a
GA-search. This leads to a new MEC, characterized by a lower
$G^\star$-value, a density $\varrho^\star$, and thus by a new straight
line of slope $1/\varrho^\star$. Intersecting this line with the
previous lines and iterating this strategy, the entire pressure-range
can be investigated in an extremely efficient way. From the economic
point of view this systematic procedure is very attractive; but it has
two additional advantages: (i) there is no risk to 'forget' MECs which
easily occurs when working on a finite $P^\star$-grid; and (ii) it
avoids a characteristic drawback of GAs that tend to fail in the
vicinity of state points that are characterized by competing
structures; in the present approach these transition points are
determined {\it exactly} via the intersection of two straight lines.

At each of the intersection points, 1000 to 3000 independent GA-runs
with 700 generations, each consisting of 500 individuals, have been
performed. Up to twelve particles per unit cell were considered, a
number which offers the system sufficient possibilities to form even
highly complex structures. Bearing all this in mind, we are confident
that the sequence of MECs which we present in the following are
complete.
 
We have considered three values of $\lambda$, corresponding to small
($\lambda = 1.5$), intermediate ($\lambda = 4.5$), and long ($\lambda
= 10$) shoulder width.  For the first case, $G^\star$ and $U^\star$,
as functions of $P^\star$, are presented in Fig. \ref{fig:thd_1.5}.
$G^\star$ is, as mentioned above, a sequence of straight lines, while
the energy levels of $U^\star$ are rational numbers, given by the
number of overlapping coronas per particle. For $\lambda = 1.5$, seven
MECs can be identified (cf. acronyms in Fig. \ref{fig:thd_1.5}). At
low pressure a columnar structure is identified, while for larger
$P^\star$-values, the relatively short corona width allows only for
compact structures.

\begin{figure}
      \begin{center}
      \begin{minipage}[t]{8.5cm}
      \includegraphics[width=8.4cm, clip] {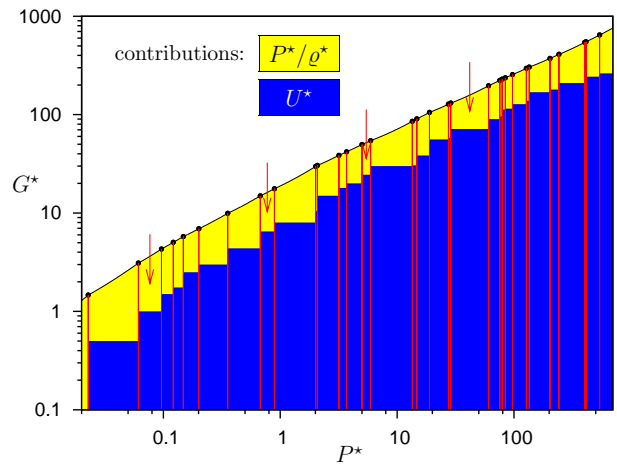}
      \end{minipage}
      \end{center}
\caption{(color online). $G^\star$ and $U^\star$ as functions of
  $P^\star$ on a double-logarithmic scale for the square-shoulder
  system, $\lambda = 4.5$. Shaded areas represent contributions to
  $G^\star$ as labeled. Arrows cf. Fig. \ref{fig:str_4.5}.}
\label{fig:thd_4.5}
\end{figure}

\begin{figure}
      \begin{center}
      \begin{minipage}[t]{8.5cm}
      \includegraphics[width=8.4cm, clip] {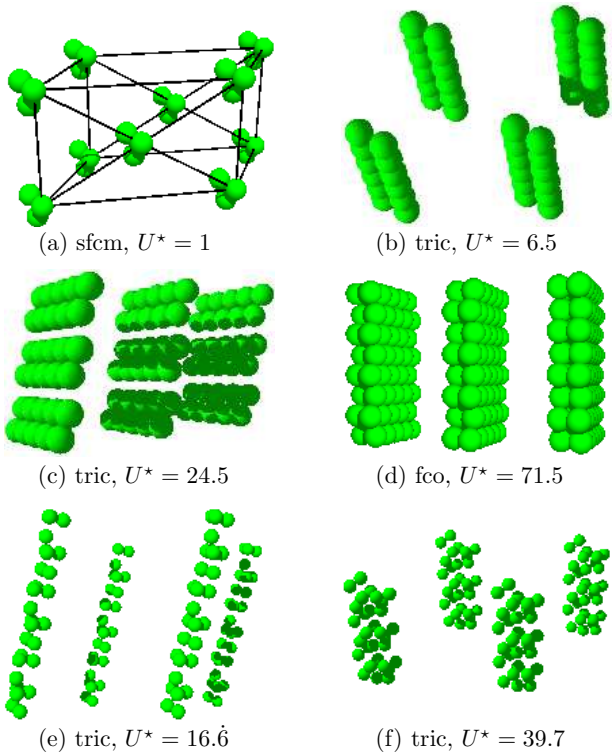}
      \end{minipage}
      \end{center}
\caption{(color online). Selected MECs for the square-shoulder system
  with $\lambda = 4.5$ [from (a) to (d)] and with $\lambda = 10$ [from
  (e) and (f)]. MECS (a) to (d) correspond to pressure values
  indicated by arrows in Fig. \ref{fig:thd_4.5}.}
\label{fig:str_4.5}
\end{figure}

\begin{figure}
      \begin{center}
      \begin{minipage}[t]{8.5cm}
      \includegraphics[width=8.4cm, clip] {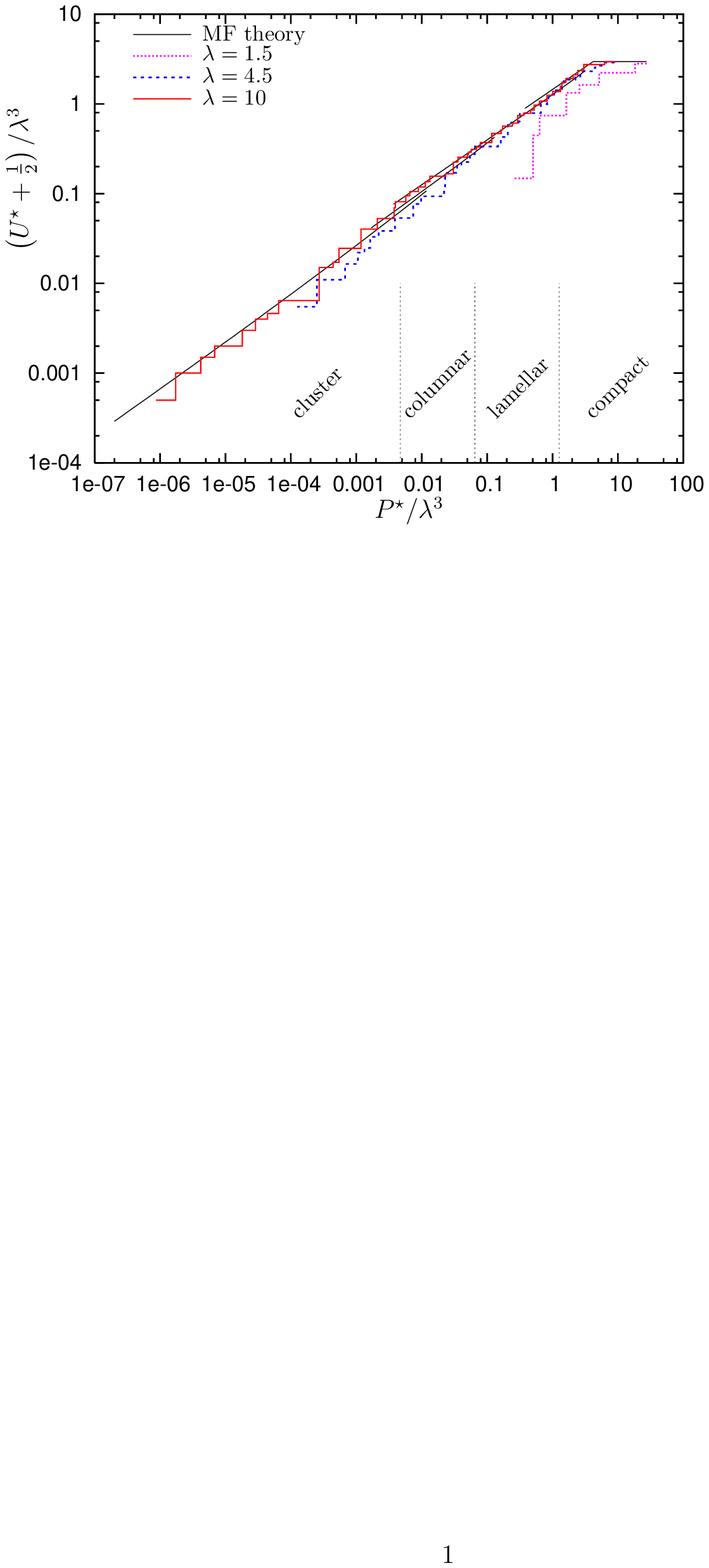}
      \end{minipage}
      \end{center}
\caption{(color online). Scaled energy per particle
  $(U^\star+1/2)/\lambda^3$ as obtained via the MF-theory (full black
  line) and for the three square-shoulder systems investigated (as
  labeled) as a function of $P^\star/\lambda^3$. The vertical, dotted
  lines indicate the borders of the four different regimes.}
\label{fig:thd_all}
\end{figure}

For $\lambda = 4.5$, the considerably broader corona offers the system
a large variety of strategies to form MECs. Their total number amounts
to 33 (cf.\ Fig. \ref{fig:thd_4.5}) which can be grouped with
increasing pressure into four structural archetypes: cluster phases,
columnar structures, lamellar phases, and, finally, compact structures
(see also Fig. \ref{fig:thd_all}). At low pressure the system forms
cluster crystals, i.e., periodic structures where the lattice points
are populated by clusters of particles. A closer analysis reveals a
strong interplay between the cluster shape and the cell geometry in
the sense that overlaps of coronas of neighboring clusters are
avoided. Inside the clusters, which contain up to eight particles, the
cores are in direct contact. For demonstration we have depicted in
Fig. \ref{fig:str_4.5}(a) a typical cluster crystal: the equilateral
triangle-shaped clusters populate the lattice positions of a single
face centered monoclinic structure. As the pressure is increased the
system develops a new strategy to minimize $G$ by forming columnar
structures. Particles aligned in close or direct contact form single
or double stranded columns, which are arranged periodically in
space. The inter-columnar spacing is imposed via the width of the
corona. An example for the columnar phase is given in Fig.
\ref{fig:str_4.5}(b). As the pressure increases further, the columns
are squeezed together in one direction, forming thereby lamellae; the
transition scenario is depicted in Fig. \ref{fig:str_4.5}(c). The
lamellar structures are realized in a large variety of
morphologies. In Fig. \ref{fig:str_4.5}(d) we show an example for a
lamellar phase: here, two hexagonally close-packed, planar layers are
in direct contact and form double layers. Finally, under the influence
of the increasing pressure, the lamellae merge, creating in this way
typical compact structures.

For $\lambda = 10$, the variety of structures is even richer, in total
48 MECs have been identified. The clusters contain more particles and
the columns are more complex in their morphology. For demonstration we
depict two of these structures in Figs. \ref{fig:str_4.5}(e) and
(f). However, the sequence of structural archetypes that has already
been identified for $\lambda = 4.5$ is maintained.

The above mentioned transition from clusters to compact structures can
be understood on a simple, semi-quantitative level. Generalizing the
ideas proposed by Glaser et al. \cite{Gla07} for two dimensional
systems, we only consider {\it aggregates} (i.e., clusters, columns,
and lamellae) instead of the individual particles. Assuming an
idealized shape for these aggregates (spheres, straight lines, and
planes) the inter- and intra-aggregate energy can be calculated. The
respective expressions are in some cases rather complex and will be
presented elsewhere \cite{Pau08}. $G^\star$ is then characterized by
$\varrho^\star$, the distance between two aggregates, and their
spatial extent. Retaining parameters up to first order and minimizing
$G^\star$ with respect to these quantities we obtain the thermodynamic
properties of the respective phase.  In particular, $(U^\star +
1/2)/\lambda^3$ as a function of $P^\star/\lambda^3$ turns out to be
$\lambda$-independent and has been plotted, along with the respective
results obtained via the GA, for all $\lambda$-values in Fig.
\ref{fig:thd_all}.  As expected, agreement improves considerably with
increasing $\lambda$. A detailed analysis of the MECs identified by
the GA (in particular for the larger $\lambda$'s) reveals, that the
MECs populate, according to their aggregate-type, nearly exclusively
the respective pressure-range.

\begin{figure}
      \begin{center}
      \begin{minipage}[t]{8.5cm}
      \includegraphics[width=8.4cm, clip] {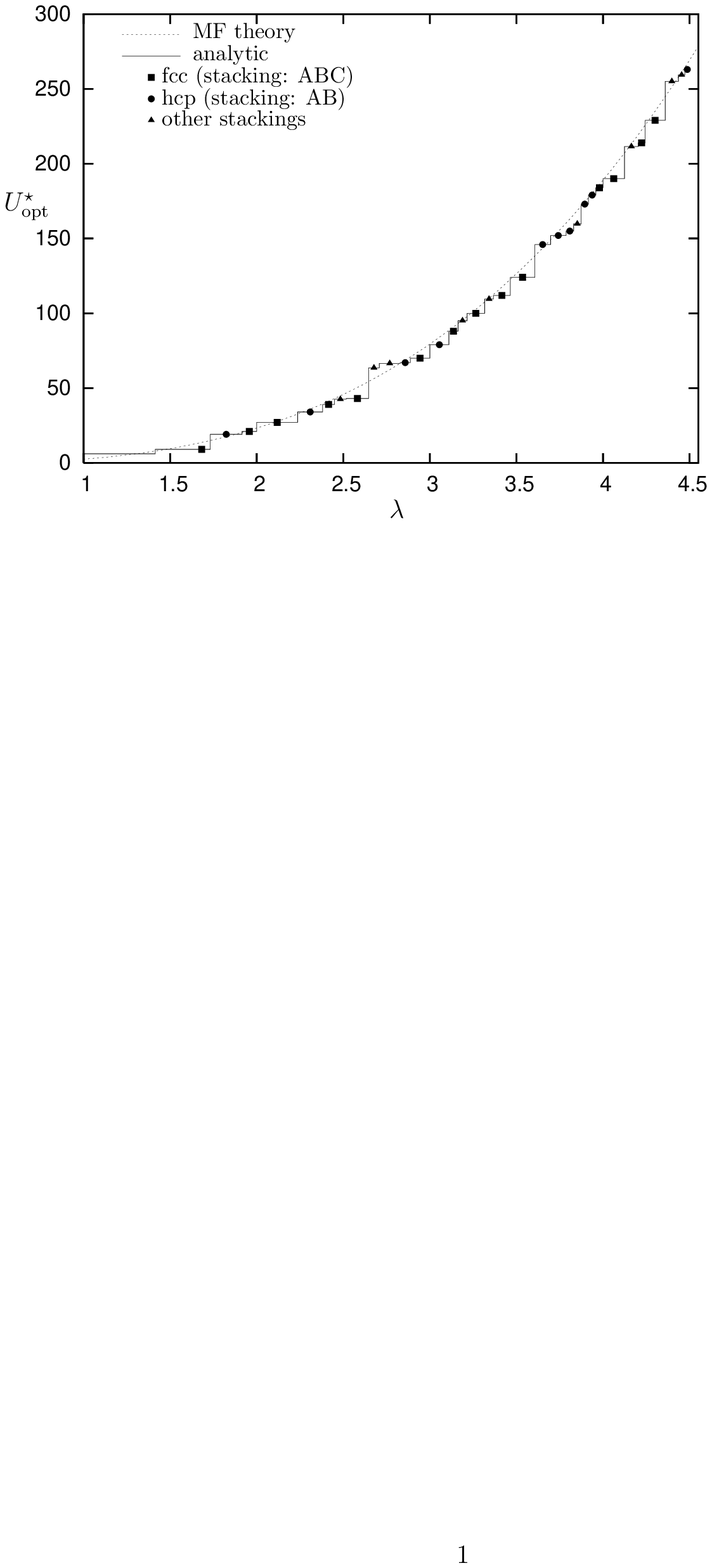}
      \end{minipage}
      \end{center}
\caption{Minimum number of corona overlaps per particle
  ($U^\star_\mathrm{opt}$) and corresponding MECs (as labeled) for
  close-packed particle arrangements as function of $\lambda$, as
  obtained via analytical considerations and the MF-theory (as
  labeled). The latter gives $U^\star(\varrho^\star=\sqrt 2) =(4\pi
  \lambda^3/3\sqrt 2)-(1/2)$.}
\label{fig:closepacked}
\end{figure}

Finally, a nice by-product should be mentioned: the well-defined range
of the shoulder represents a highly sensitive antenna to discern
between competing structures at {\it close-packed}
arrangements. Varying $\lambda$ from 1 to 4.5, we have identified via
analytic considerations not only the usual suspects for close-packed
scenarios, namely hcp and fcc; also other, more complicated stacking
sequences are observed for particular $\lambda$-values --
cf. Fig. \ref{fig:closepacked}.

Summarizing, our investigations have given {\it quantitative} evidence
about the rich wealth of self-assembly scenarios of soft particles
that interact via a simple, radially symmetric pair potential: while
at high pressure compact structures are dominant, we observe at
intermediate and low pressure low-symmetry structures, that include
clusters, columns, and lamellae. The simple functional form allows to
{\it understand} fully analytically the system's strategies to
self-organize in such complex scenarios. Our findings can also be of
technological relevance: since our GA-based search strategy is
reliable and efficient, it can easily be applied to systems with more
complex interactions, pointing thus towards more technological
applications, such as nano-lithography or nano-electricity.

The authors are indebted to Dieter Gottwald and Julia Fornleitner
(both Wien) for stimulating discussions. Financial support by the
Austrian Science Foundation under Proj.~Nos.~W004, P17823-N08, and
P19890-N16 is gratefully acknowledged.


\begin{thebibliography}{10}
\bibitem{comment} For two dimensional simple lattices the parameter
space is two-dimensional. Here, the region that corresponds to
unphysical, i.e., overlapping particle configurations can be
identified rather easily: it is given by infinitely many circles which
become for $\varrho \sigma^2$ the so-called Ford circles; the latter
represent a super-set of the Appolonian Gasket, i.e., an object of
fractal dimension of 1.3057.
\end{thebibliography}

\end{document}